\documentclass[transmag,journal]{IEEEtran}
\usepackage[T1]{fontenc}
\usepackage[latin9]{inputenc}
\usepackage{amsmath}
\usepackage{graphicx}

\usepackage{fancyhdr}
 
\usepackage[unicode=true,
 bookmarks=true,bookmarksnumbered=true,bookmarksopen=true,bookmarksopenlevel=1,
 breaklinks=false,pdfborder={0 0 0},pdfborderstyle={},backref=false,colorlinks=false]
 {hyperref}
\hypersetup{pdftitle={Your Title},
 pdfauthor={Your Name},
 pdfpagelayout=OneColumn, pdfnewwindow=true, pdfstartview=XYZ, plainpages=false}

\makeatletter

\newcommand*\LyXZeroWidthSpace{\hspace{0pt}}

\let\oldforeign@language\foreign@language
\DeclareRobustCommand{\foreign@language}[1]{%
  \lowercase{\oldforeign@language{#1}}}

\usepackage[caption=false,font=normalsize,labelfont=sf,textfont=sf]{subfig}

\@ifundefined{showcaptionsetup}{}{%
 \PassOptionsToPackage{caption=false}{subfig}}
\usepackage{subfig}
\makeatother

\begin{document}

\title{Exceptional Degeneracy in a Waveguide Periodically Loaded with Discrete
Gain and Radiation Loss Elements}
\author{\IEEEauthorblockN{Ahmed~F.~Abdelshafy, Tarek~Mealy, Ehsan~Hafezi, Alireza~Nikzamir
and Filippo~Capolino}\IEEEauthorblockA{Electrical Engineering and Computer Science Department,\\ University of California, Irvine, CA 92697 USA}}


\IEEEtitleabstractindextext{

\begin{abstract}
We demonstrate that a periodic waveguide comprising of uniform lossless
segments together with discrete gain and radiating elements supports
exceptional points of degeneracy (EPDs). We provide analytical expressions
for all possible conditions that guarantee the occurrence of an EPD,
i.e., the coalescence of eigenvalues and eigenvectors. We show that
EPDs are not only achieved using symmetric gain and radiation periodic
loading, but they are also obtained using asymmetric gain and radiation
loss conditions. We illustrate the characteristics of the degenerate
electromagnetic modes, showing the dispersion diagram and discussing
the tunability of the EPD frequency. We show a special condition,
we refer to it as parity-time (PT)-glide symmetry, which leads to
a degeneracy that is occurring at all frequencies of operation. The
class of EPDs proposed in this work is very promising for many applications
that incorporate discrete-distributed coherent sources and radiation-loss
elements; operating in the vicinity of such special degeneracy conditions
leads to potential performance enhancement in a variety of microwave
and optical resonators, antennas, and devices and can be extended
to a new class of active integrated antenna arrays and radiating laser
arrays.
\end{abstract}


}

\maketitle


\thispagestyle{fancy}

\IEEEdisplaynontitleabstractindextext{}

\IEEEpeerreviewmaketitle{}

Electromagnetic (EM) guiding structures or resonators are described
by their eigenmodes' (eigenvalues and eigenvectors) evolution equations.
Eigenmodes representing EM propagating waves in a multimodal waveguide
may coalesce into a single degenerate eigenmode by varying at least
one parameter of the parameter space (frequency, geometrical/physical
parameters) of the waveguide system; this special point in the system
parameter space is called an exceptional point of degeneracy (EPD)
\cite{kato_perturbation_1995,heiss_physics_2012}. At the EPD, two
or more eigenstates of the system coalesce into a single degenerate
eigenstate. Such condition is simply referred to as 'EP' in various
works; here the \textquoteleft D\textquoteright{} is used to stress
the importance of degeneracy \cite{berry_physics_2004}. The number
of degenerated eigenstates is referred to as the order of the exceptional
point. In the proximity of an EPD angular frequency $\omega$, the
eigenvalues $\lambda$ associated with the coalescing eigenvectors
change with respect to frequency as $(\omega-\omega_{e})\propto(\lambda-\lambda_{e})^{n}$,
in which $\lambda_{e}\,,\omega_{e}$, and $n$ are the degenerate
eigenvalue, EPD angular frequency, and order of EPD, respectively.

In general, an EPD occurs in a system where the space-time evolution
of the system vector is characterized by a non-Hermitian matrix, which
can be imposed also by periodicity in space \cite{figotin_gigantic_2005,othman_experimental_2017,Figotin2007Slow,abdelshafy_exceptional_2019}
or in time \cite{kazemi_exceptional_2019,rouhi_exceptional_2020}
or by having losses and gain in the system \cite{othman_theory_2017,abdelshafy_exceptional_2019}
including systems satisfying parity-time (PT) symmetry \cite{heiss_physics_2012,ruter_observation_2010,guo_observation_2009}.
The unique degenerate dispersion behavior is accompanied by supreme
characteristics including the vanishing of the group velocity \cite{figotin_frozen_2006,gutman_slow_2012}
as well as the dramatic improvement in the local density of states
\cite{othman_giant_2016} resulting in a robust increase in the loaded
quality factor of the structure. The EPD phenomenon has been proved
to have various applications including high quality factor ($Q$)
and low-threshold lasers \cite{veysi_degenerate_2018}, lasers in
coupled ring resonators\cite{hodaei_parity-timesymmetric_2014}, and
low-threshold oscillators \cite{othman_low_2016,abdelshafy_electron-beam-driven_2018,abdelshafy_distributed_2020}.
Moreover, the deviation of the perturbed eigenvalues from the degenerate
eigenvalue is large when a small perturbation to a system parameter
is applied; so this sensitivity brings another class of applications
in sensors \cite{wiersig_sensors_2016,hodaei_enhanced_2017,ren_ultrasensitive_2017,kazemi_ultra-sensitive_2020}.

\begin{figure}[t]
\begin{centering}
\subfloat[]{\begin{centering}
\includegraphics[width=0.85\columnwidth]{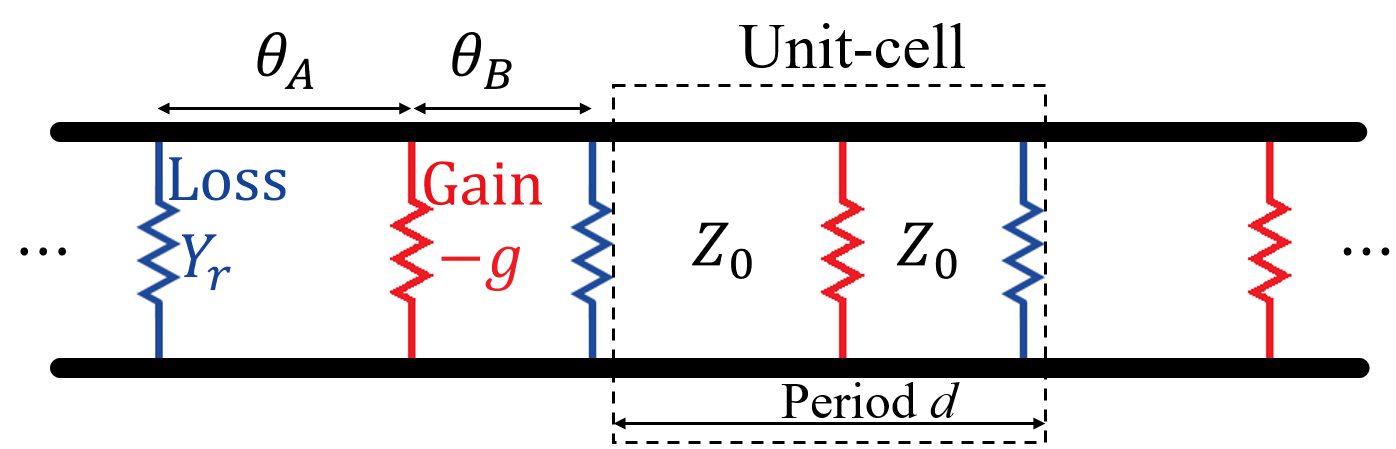}
\par\end{centering}
}
\par\end{centering}
\begin{centering}
\subfloat[]{\begin{centering}
\includegraphics[width=0.75\columnwidth]{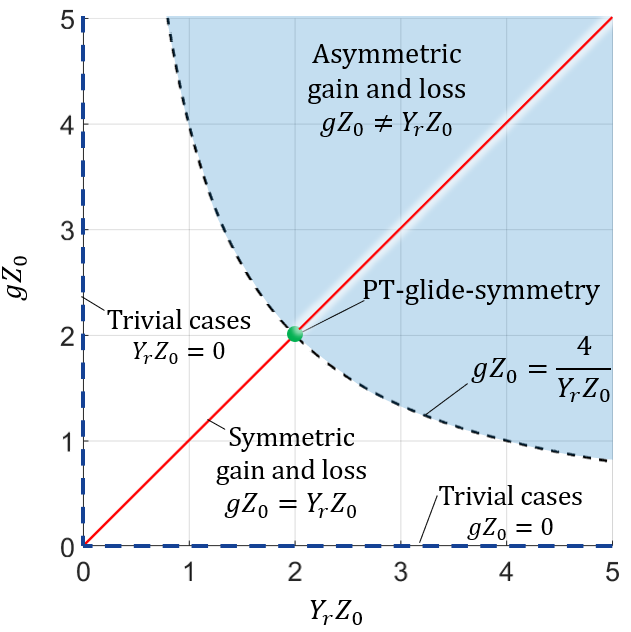}
\par\end{centering}
}
\par\end{centering}
\centering{}\caption{\label{fig:Example-geometry-1-1} (a) Unit cell schematic of a periodic
waveguide, represented by its equivalent transmission line (TL), made
of two segments with characteristic impedance $Z_{0}$ and loaded
with shunt lossy element ($Y_{r}$) and shunt gain element ($-g$).
(b) The relation between gain and loss to have an EPD. The shaded
area represents the asymmetric cases where gain and loss relation
to have an EPD is $gZ_{0}=4/(Y_{r}Z_{0}\sin^{2}(\theta_{A}))$. The
black-dashed curve represents one of the asymmetric cases when $\theta_{A}=\theta_{B}=(2m+1)\pi/2$,
also the PT-glide-symmetry case is depicted by the green dot in the
intersection between the asymmetric case dashed curve and red line
representing symmetric cases.}
\end{figure}

In this paper, we present an example of a waveguide that exhibits
a second-order EPD by periodically loading a uniform waveguide with
gain and radiating elements, as schematized in Fig. \ref{fig:Example-geometry-1-1}.
We provide the analytical expressions for the second-order EPD conditions
to occur in different loading configurations for the gain and radiating
elements. The EPD condition is observed in the dispersion diagram
and by the coalescence of the eigenvectors. We also describe the Floquet-Bloch
impedance in the vicinity of the EPD, which can be important for matching
and stability analysis. We conclude by showing a possible application
as an array of radiation elements oscillating and radiating at the
EPD frequency.

We consider a uniform waveguide that is periodically loaded with discrete
gain and loss and that is schematically represented by its equivalent
transmission line model (TL) \cite{marcuvitz_representation_1951,felsen_radiation_1994};
this model can be applied to waveguides operating from microwaves
to optics, hence our formulation is general. We assume that the waveguide
is periodically loaded with discrete shunt gain and resistive elements,
as shown in Fig. \ref{fig:Example-geometry-1-1}(a). Indeed, it is
customary to represent radiation from discrete points along a waveguide
using resistive loads.

The periodic unit cell is divided into four parts: two uniform waveguide
segments together with a discrete gain element and a discrete radiative
element represented by its equivalent resistance. For simplicity the
waveguide segments are assumed to have similar characteristic impedance,
but with possibly different electrical lengths $\theta_{A}=k_{0}l_{A}$,
and $\theta_{B}=k_{0}l_{B}$ where $l_{A}$ and $l_{B}$ are the physical
lengths of the waveguide segments A and B, respectively, and $k_{0}=\omega/v_{ph}$
is the waveguide propagation constant, with $v_{ph}$ being the phase
velocity of a uniform-waveguide mode. It is convenient to define a
system state vector as $\boldsymbol{\Psi}(z)=[\begin{array}{cc}
V(z), & I(z)\end{array}]^{T}$ , with $T$ indicating the transpose action. Therefore, referring
to Fig. \ref{fig:Example-geometry-1-1}(a), we use the transfer matrix
of a shunt element $\underline{\mathbf{T}}_{shunt}$ and lossless
transmission line $\underline{\mathbf{T}}_{TL}$ \cite{pozar_microwave_2009},
and we form a relation between equivalent voltage and current between
the two ends of a unit cell as $\boldsymbol{\Psi}_{n+1}=\underline{\mathbf{T}}_{\textrm{U}}\boldsymbol{\Psi}_{n}$.
The unit cell transfer matrix $\underline{\mathbf{T}}_{\textrm{U}}$
is the result of the multiplication of four transfer matrices as
\begin{equation}
\underline{\mathbf{T}}_{\textrm{U}}=\underline{\mathbf{T}}_{shunt}(Y_{r})\underline{\mathbf{T}}_{TL}(\theta_{B})\underline{\mathbf{T}}_{shunt}(-g)\underline{\mathbf{T}}_{TL}(\theta_{A}).
\end{equation}

We look for solutions of the type $\boldsymbol{\Psi}_{n}\propto\boldsymbol{\Psi}_{0}e^{-jknd}$
satisfying the Floquet\textquoteright s condition $\boldsymbol{\Psi}_{n+1}=e^{-jkd}\boldsymbol{\Psi}_{n}$,
where $d$ is the waveguide period, $k$ is the Floquet-Bloch wavenumber,
and we implicitly assume the time convention $e^{j\omega t}$. Hence,
the eigenmodes supported in such a system are described by the eigenvalue
problem 
\begin{equation}
\left[\underline{\mathbf{T}}_{\textrm{U}}-\lambda\underline{\mathbf{I}}\right]\boldsymbol{\Psi}=\mathbf{0},\label{eq:Eigenvalue prob}
\end{equation}
where $\underline{\mathbf{I}}$ is the identity matrix of order two,
$\lambda=e^{-jkd}$ is an eigenvalue and $\boldsymbol{\Psi}$ is the
associated eigenvector. The eigenvalues are readily found by solving
the characteristic equation $\det(\underline{\mathbf{T}}_{\textrm{U}}-\lambda\underline{\mathbf{I}})=0$,
i.e., by finding the roots of the characteristic polynomial

\begin{equation}
\begin{array}{c}
\lambda^{2}+\Big[-2\cos(\theta_{A}+\theta_{B})-gY_{r}Z_{0}^{2}\sin(\theta_{A})\sin(\theta_{B})\ \ \ \ \ \ \ \ \\
\ \ \ \ \ \ \ \ \ \ \ \ \ -jZ_{0}Y_{r}(1-g/Y_{r})\sin(\theta_{A}+\theta_{B})\Big]\lambda+1=0,
\end{array}\label{eq:C/chs Poly.}
\end{equation}
where $Z_{0}$ is the characteristic impedance of the two uniform
waveguide segments. To have two identical roots in a second-order
polynomial of the form of $\lambda^{2}+a\lambda+b=0$, $a^{2}-4b$
must vanish. Having $b=1$ in the proposed system characteristic polynomial
(\ref{eq:C/chs Poly.}) indicates that the eigenvalues are $\lambda_{1}=1/\lambda_{2}=e^{-jk_{1}d}$,
which implies that $k_{1}=-k_{2}$. Also, the necessary and sufficient
conditions for having identical eigenvalues $\lambda$ is $a=(-1)^{p}2;$
where $p$ is an integer number indicating the positive and negative
possible solutions of $a$. Therefore conditions that must be satisfied
at the EPD frequency, related to the real and imaginary parts of $a$,
read as

\begin{equation}
-2\cos(\theta_{A}+\theta_{B})-gY_{r}Z_{0}^{2}\sin(\theta_{A})\sin(\theta_{B})=(-1)^{p}2\label{eq:realnece}
\end{equation}

\begin{equation}
Z_{0}Y_{r}(1-g/Y_{r})\sin(\theta_{A}+\theta_{B})=0.\label{eq:imag.}
\end{equation}

The second condition, (\ref{eq:imag.}), is satisfied either by constraining
the gain and radiation-element equivalent resistance (i.e., $1-g/Y_{r}=0$)
or by constraining the TL segments' electrical lengths (i.e., $\sin(\theta_{A}+\theta_{B})=0$).
Whereas, the first condition in (\ref{eq:realnece}) is used as the
design equation for different possible cases that are leading to identical
eigenvalues. The chart in Fig. \ref{fig:Example-geometry-1-1}(b)
summarizes the required relation between gain and radiation loss to
have an EPD, that are discussed next.

\paragraph{Trivial cases with vanishing gain or loss ($Y_{r}=0$ or $g=0)$}

A trivial condition to satisfy EPD is by having $\theta_{A}+\theta_{B}=p\pi,$
where $p$ is an integer, besides having either $g=0$ or $Y_{r}=0$,
represented by the blue-dashed vertical or horizontal line, respectively,
in the chart in Fig. \ref{fig:Example-geometry-1-1}(b). The EPD obtained
for this case occurs at $k=\pi/d$, where $d$ is the unit cell period.
However, we do not focus on this trivial case as it is not suitable
for applications that incorporate \textit{both} discrete-distributed
coherent sources and radiation loss elements.

\paragraph{Symmetric gain and loss cases $(Y_{r}=g)$}

One possibility to satisfy the condition in (\ref{eq:imag.}), $\mathrm{Im}(a)=0$,
is by enforcing balanced gain and radiation-loss, $g=Y_{r}$ represented
by the solid-red line in the chart in Fig. \ref{fig:Example-geometry-1-1}(b).
The normalized gain and radiation elements values that satisfy the
other EPD condition (\ref{eq:realnece}) $\mathrm{Re}(a)=2(-1)^{p}$,
yields

\begin{equation}
Y_{r}Z_{0}=gZ_{0}=\sqrt{\dfrac{2\left((-1)^{p}-\cos(\theta_{A}+\theta_{B})\right)}{\sin(\theta_{A})\sin(\theta_{B})}}.\label{eq:symmetrci EPD cond}
\end{equation}

It is important to mention that for any arbitrary choice of $\theta_{A}$
and $\theta_{B},$ the term inside the root in (\ref{eq:symmetrci EPD cond})
can always have a positive value by choosing the proper $p$ value.
The symmetrical gain and radiation-loss is a straightforward condition
that leads to EPD where the introduced amount of radiation loss should
be compensated with the same amount of gain in order to have neither
a decaying nor a growing wave. Although, in this case, gain and loss
loads are equal, the unit cell with $\theta_{A}\neq\theta_{B}$ does
not classify as (PT)-symmetric condition, that could be defined based
on the system\textquoteright s refractive index obeying $n(z)=n^{*}(-z)$,
where $z$ is a coordinate in the system and $*$ denotes complex
conjugation \cite{el-ganainy_theory_2007,ruter_observation_2010,heiss_physics_2012}.
Enforcing spatial symmetry in the unit cell by choosing equal electrical
lengths $\theta_{A}=\theta_{B}$ leads to a unit cell that satisfies
a possible definition of PT-symmetry as used in \cite{wei_scattering_2019}.
Indeed, the symmetric load case with $\theta_{A}=\theta_{B}$ satisfies
the following

\begin{equation}
n(z+\frac{d}{2})=n^{*}(z),\label{eq:pt- glide symmtry case}
\end{equation}
which holds the reflection between gain and loss by the complex conjugate
operator $*$ and the translation along $z$ by half a period. We
define the PT-glide-symmetry condition as in (\ref{eq:pt- glide symmtry case})
which can be used also to describe more complicated structures. In
general, glide symmetry is a symmetry operation comprised of a reflection
operation over a certain coordinate and translation along with that
coordinate \cite{hessel1973propagation,bagheriasl2019bloch,Mock:20}.
We refer to the condition in (\ref{eq:pt- glide symmtry case}) as
PT-glide-symmetry where the EPD condition met at every frequency with
$\lambda_{e}=e^{-jk_{e}d}=1$. Note that for the reciprocal system
we are studying, the EPD is only possible when $k_{e}$ is purely
real with the value of either $k_{e}d=0$ or $k_{e}d=\pi$.

\begin{figure}[t]
\centering{}\centering\includegraphics[width=0.75\columnwidth]{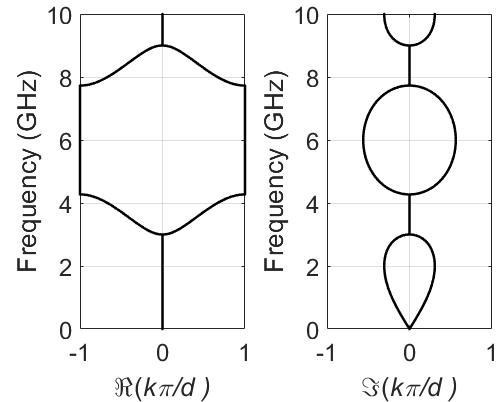}\caption{\label{fig:dispersion_sym}Dispersion diagram of complex-valued wavenumber
versus frequency for gain-radiation loss symmetric case, $Y_{r}=g=2\sqrt{2}/Z_{0}=\sqrt{2}/25$~S,
with $3\theta_{A}=\theta_{B}=3\pi/4$ at 3 GHz. Wavenumber degeneracies
are observed at 3 GHz, 4.333 GHz, 7.715 GHz, 9 GHz, etc., where either
$k=0$ or $kd/\pi=\pm1$.}
\end{figure}

Figure \ref{fig:dispersion_sym} depicts the dispersion of a waveguide
with symmetric loads and different electrical lengths. The waveguide
exhibits EPD by satisfying the EPD condition in (\ref{eq:symmetrci EPD cond})
at $3$ GHz such that $3\theta_{A}=\theta_{B}=\pi/4$, $Z_{0}=50\,\Omega$,
and $Y_{r}=g=2\sqrt{2}/Z_{0}=\sqrt{2}/25$~S.

In the rest of the paper, we focus on the case with asymmetric gain
and radiation-loss as it provides more flexibility in using different
values of gain and radiation-loss. Indeed, the value of resistance
of the radiating element cannot be set arbitrarily and the constraints
depend on the specific design, while gain usually can be tuned by
simply changing a biasing voltage.

\paragraph{Asymmetric gain and loss cases $(Y_{r}\protect\neq g)$}

Condition (\ref{eq:imag.}) can be satisfied also for asymmetric gain
and radiation-loss cases, represented by the shaded area in the chart
in Fig. \ref{fig:Example-geometry-1-1}(b), by constraining the waveguide
segments' electrical lengths as

\begin{equation}
\theta_{A}+\theta_{B}=p\pi,\label{eq:EPD asym length}
\end{equation}
where $p$ is an integer number, in other words, the total length
of the waveguide of a unit cell at $f_{e}$ is an integer multiple
of half wavelength. The other condition (\ref{eq:realnece}) forces
the relation between the normalized gain and radiation-loss values
to be 
\begin{equation}
\left(gZ_{0}\right)=\dfrac{4}{\left(Y_{r}Z_{0}\right)\sin^{2}(\theta_{A})}.\label{eq:EPD asym impedance}
\end{equation}

\begin{figure}[t]
\begin{centering}
\subfloat[]{\begin{centering}
\centering\includegraphics[width=0.65\columnwidth]{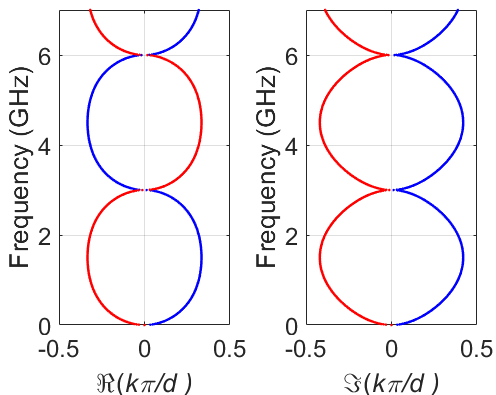}
\par\end{centering}
}
\par\end{centering}
\centering{}\LyXZeroWidthSpace\subfloat[]{\begin{centering}
\centering\includegraphics[width=0.65\columnwidth]{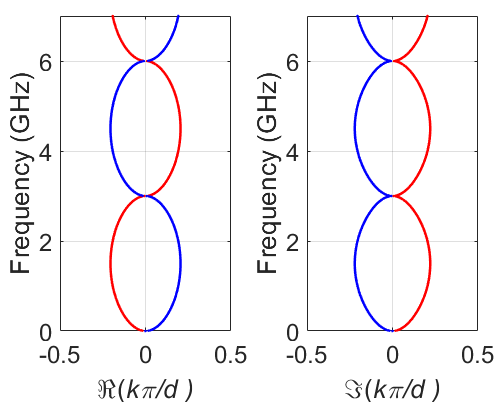}
\par\end{centering}
}\caption{\label{fig:dispersion_first_example}Dispersion diagram of complex-valued
wavenumber versus frequency. Wavenumber degeneracies are observed
at 3 GHz, 6 GHz, etc. where both wavenumbers vanish. The two wavenumbers
are denoted by different colors, for the two different cases with
$Z_{0}=50\,\Omega$ and $\theta_{A}=\theta_{B}=\pi/2$ at 3 GHz: (a)
$Y_{r}=20$ mS, $g=80\,$mS corresponding to $Y_{r}Z_{0}<2$, and
(b) $Y_{r}=50$ mS, $g=32\,$mS corresponding to $Y_{r}Z_{0}>2$}
\end{figure}

By forcing these two conditions (\ref{eq:EPD asym length},\ref{eq:EPD asym impedance}),
the degenerate eigenvalue of the eigenvalue problem in (\ref{eq:Eigenvalue prob})
is equal to
\begin{equation}
\lambda_{e}=e^{-jk_{e}d}=\begin{cases}
(-1)^{p+1}, & \textrm{if}\,\theta_{A}\neq\theta_{B}\neq l\pi\\
(-1)^{p}, & \textrm{otherwise}
\end{cases}
\end{equation}
where $l$ is an integer such that $0\prec l\prec m$, and the degenerate
eigenvector is $\boldsymbol{\Psi}_{e}=I\left[\begin{array}{cc}
Z_{B,e}, & 1\end{array}\right]^{T}$ , where 
\begin{equation}
Z_{B,e}=-2Z_{0}/\left(Y_{r}Z_{0}+j2\cot\theta_{A}\right),
\end{equation}
is the Bloch impedance of the degenerate mode. Figure \ref{fig:dispersion_first_example}(a)
depicts the dispersion of one waveguide that exhibits EPD by satisfying
the conditions at $3$ GHz such that $\theta_{A}=\theta_{B}=\pi/2$,
$Y_{r}=20$~mS, $Z_{0}=50\,\Omega$, and $g=80\,$mS to satisfy the
EPD condition in (\ref{eq:EPD asym impedance}). The two complex wavenumbers
are traced in two different colors such that one can observe the coalescence
of the two complex wavenumbers at the EPD frequency and its harmonics
(i.e. all meet the EPD conditions). Note that the EPD points are the
transition points at which the complex wavenumbers alternate between
the same sign for both parties of the complex wavenumber (i.e. real
and imaginary parts) indicating growing waves to opposite signs indicating
decaying waves.

Upon analyzing the modal dispersion equation, it can be proved that
when the special case of $Y_{r}Z_{0}=2$, accordingly, $g=Y_{r}$
(i.e., symmetric case) and $\theta_{A}=\theta_{B}$ are met utilizing
the aforementioned PT-glide-symmetry case, then the two eigenvalues
(and also the eigenvectors) will be identical at every frequency.
In Fig. \ref{fig:dispersion_first_example}(a) we show an example
of the dispersion when $\theta_{A}=\theta_{B}$ when $Y_{r}Z_{0}<2$,
that is true when using the aforementioned parameters, whereas in
Fig. \ref{fig:dispersion_first_example}(b) we show an analogous example
that exhibits EPD at the same frequency when the condition reads as
$Y_{r}Z_{0}>2$ by selecting $Y_{r}=$ 50 mS and $g=32$~mS for the
same $Z_{0}=50\,\Omega$.

The periodic electromagnetic guiding structure is characterized by
the modal dispersion equations (\ref{eq:C/chs Poly.}). Each eigenmode
is characterized by its eigenvalue $\lambda_{i}=e^{-jk_{i}d}$ related
to the associated complex Floquet wavenumbers $k_{i}$ and its eigenvector
$\boldsymbol{\Psi}_{i}=I_{i}\left[\begin{array}{cc}
Z_{B,i}, & 1\end{array}\right]^{T}$, with $i=1,2$ for the case under study here, where $Z_{B,i}$ is
the $i^{th}$ mode Floquet-Bloch impedance.

\begin{figure}[t]
\begin{centering}
\centering
\par\end{centering}
\centering{}\includegraphics[width=1\columnwidth]{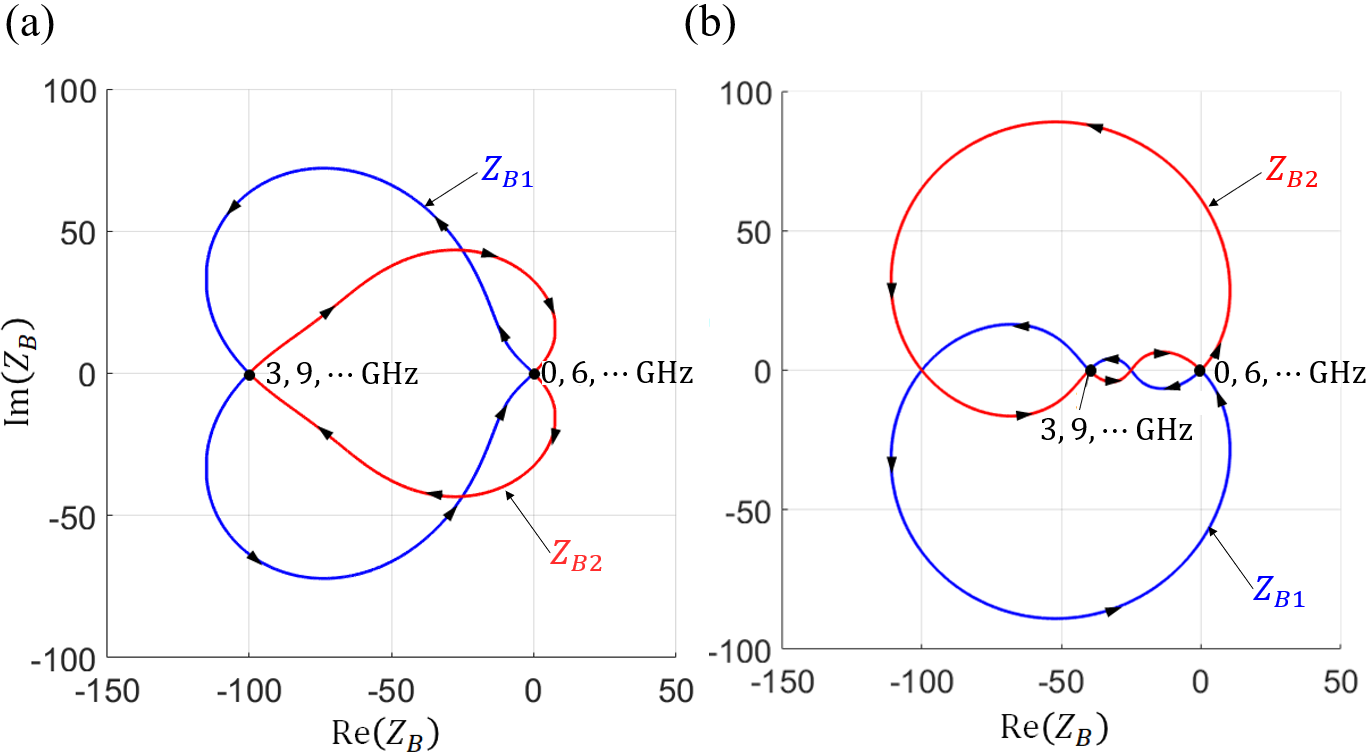}\caption{\label{fig:ZB_first_example}Complex-valued Bloch impedances $Z_{B}$
showing the trajectory of $Z_{B}$ evolution varying frequency where
arrows represent the direction of frequency increasing. Degeneracies
are observed at 3 GHz, 6 GHz, etc. where both wavenumbers vanish,
i.e. $k=0$. The two modes' Bloch impedance are denoted by different
colors, matching different modes' colors in the dispersion diagram
in Fig. 2 (a, b), for the two different cases with $\theta_{A}=\theta_{B}=\pi/2$
at 3 GHz: (a) $Y_{r}=20$ mS corresponding to $Y_{r}Z_{0}<2$, and
(b) $Y_{r}=50$ mS corresponding to $Y_{r}Z_{0}>2$.}
\end{figure}

The evolution of the eigenmodes' complex Floquet-Bloch impedance varying
frequency, which is directly related to the evolution of the eigenvectors
$\boldsymbol{\Psi}_{i}$, is shown next. The coalescence of the eigenvectors
at the EPD is based on having a degenerate Floquet-Bloch impedance
(i.e., $Z_{B,e}=Z_{B,1}=Z_{B,2}$). Figure \ref{fig:ZB_first_example}
shows the trajectory of the complex Bloch impedance $Z_{B,i}$ for
increasing frequency for two different cases: (a) $Y_{r}Z_{0}<2$
depicted in Fig. \ref{fig:ZB_first_example}(a), and (b) $Y_{r}Z_{0}>2$
depicted in Fig. \ref{fig:ZB_first_example}(b), associated with dispersion
diagrams shown in Fig. \ref{fig:dispersion_first_example}(a,b), respectively.
It is obvious from the traces shown in Fig. \ref{fig:ZB_first_example}(a,b)
that in general, the Bloch impedances are complex conjugate to each
other over the whole frequency range except at the EPD frequency $3$
GHz and its harmonics $6,9,\ldots$ GHz where they become purely real.
At the EPD, the two impedances turn into one degenerate real impedance
$Z_{B,e}$ either $-2/Y_{r}$ or zero.

We describe succinctly some possible applications that incorporate
discrete-distributed coherent sources and radiation elements (that
are usually characterized by loss lumped elements, like the admittances
$Y_{r}$). One application to the proposed EPD scheme is an active
radiating oscillator that requires the incorporation of discrete-distributed
coherent sources and radiation loss elements. This active oscillator
is realized in a cavity made of a finite-length waveguide exhibiting
EPD with asymmetric gain and loss. As proof of the concept, and regardless
of the specific implementation, the radiating elements are simply
modeled as a distributed shunt radiation loss, whereas gain is modeled
in each unit cell using non-linear cubic $i\textrm{-}v$ characteristic
$i(t)=-gv(t)+\zeta v^{3}(t)$ of the active device \cite{oshmarin_new_2019,abdelshafy_distributed_2020}
which can be practically implemented with circuits with amplifying
devices, such as CMOS transistors or Op-Amps, with positive feedback.
Here $-g$ is the small-signal slope of the $i\textrm{-}v$ curve
in the negative resistance region, and $\zeta$ is the third-order
non-linearity constant that models the saturation characteristic of
the device. We set the turning point $v_{b}=\sqrt{g/(3\zeta)}$ of
the $i\textrm{-}v$ characteristics to be $1$ volt, and accordingly,
we set $\zeta=g/3$.

\begin{figure}[tbh]
\begin{centering}
\centering\includegraphics[width=1\columnwidth]{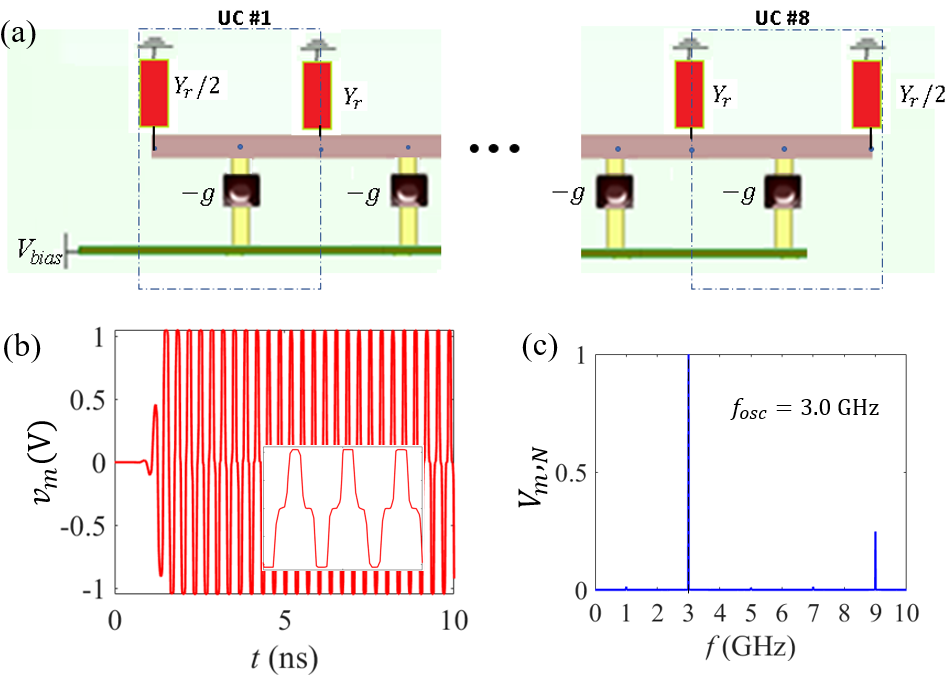}
\par\end{centering}
\caption{\label{fig:EPD-oscillator-consisting}EPD oscillator consisting of
8 cascaded unit cells (UCs) loaded with gain and loss (representing
a radiating antenna) as shown in Fig.  \ref{fig:Example-geometry-1-1}(a).
Active gain devices are placed in each UC from the TL to the bias
line (that acts as a ground for a.c. signals). (b) Voltage waveform
$v_{m}(t)$ monitored at the $Y_{r}$ load in the middle of the structure
where steady-state oscillation is observed in less than 2 ns. (c)
Normalized voltage spectrum $V_{m,N}$ ($f$) shows that oscillations
occur at around 3 GHz, that corresponds to the EPD frequency of 3
GHz in Fig.  \ref{fig:dispersion_first_example}(b).}
\end{figure}

We tested the finite-length loaded cavity comprised of 8 unit cells
as shown in Fig.  \ref{fig:EPD-oscillator-consisting}(a) in the time
domain solver implemented in Cadence Virtuoso IC 616. The unit cell
is chosen to have identical ideal TL segments with $Z_{0}=50\,\Omega$
and each has an electric length $\theta(3\textrm{GHz})=\pi/2$. The
gain and loss elements are chosen as $g=32$~mS, $Y_{r}=50\,$mS
to satisfy the EPD condition in (\ref{eq:EPD asym impedance}). Accordingly,
we report that the oscillation occurs close to the EPD frequency and
the waveform $v_{m}(t)$, at the load $Y_{r}$ in the middle of the
structure between the fourth and the fifth unit cell reaches a steady-state
in less than 2 ns as shown in Fig.  \ref{fig:EPD-oscillator-consisting}(b).
The oscillation frequency is determined by taking the Fourier transform
of $v_{m}(t)$ in the time window from 2 to 100 ns, shown in Fig.
 \ref{fig:EPD-oscillator-consisting}(c), and it confirms the oscillatory
behavior around the EPD frequency 3 GHz and its odd harmonics (9,
15, ...) GHz since they all satisfy EPD conditions. 

In summary, we have demonstrated that a periodic waveguide loaded
with gain and radiating elements as shown in Fig. \ref{fig:Example-geometry-1-1}(a)
exhibits EPDs. We have shown the different conditions for having EPDs
summarized in Fig. \ref{fig:Example-geometry-1-1}(b) and also, importantly,
we have demonstrated a case where the EPD condition is met at every
frequency satisfying the PT-glide-symmetry condition. The theoretical
framework developed applies to various structures operating from microwave
to optical frequencies. The discrete radiation admittances considered
in this paper represent the input admittances of a periodic array
of antennas. We have shown that EPDs occur at frequencies where the
two TL wavenumbers vanish, leading to possible applications of broadside
radiation in arrays of antennas periodically connected to the waveguide.
Such a phenomenon may pave the way to a new class of active traveling-wave
antennas and also in array antennas with all elements oscillating
and synchronized.



\section*{Acknowledgment}

This material is based upon work supported by the National Science
Foundation under award NSF ECCS-1711975.

\bibliographystyle{IEEEtran}
\bibliography{MyLibraryNoURL.bib}


\end{document}